\documentclass{ws-ijmpa}

\usepackage[super]{cite}
\usepackage{xcolor}
\usepackage{amsmath}
\usepackage[verbose,hypertexnames=false]{hyperref}
\hypersetup{colorlinks=false,allbordercolors=blue,pdfborderstyle={/S/U/W 1}}
\usepackage{multirow}  
\usepackage[section]{placeins}

\begin{document}

\markboth{V. Patel and A. K. Rai}{Spectroscopy of \(ncnc\bar{c}\), \(nnnn\bar{c}\), and \(nnnc\bar{c}\) pentaquarks within the framework of Regge phenomenology}

%
\catchline{}{}{}{}{}
%

\title{Spectroscopy of \(ncnc\bar{c}\), \(nnnn\bar{c}\), and \(nnnc\bar{c}\) pentaquarks within the framework of Regge phenomenology
}

\author{Vandan Patel
}

\address{Department of Physics, Sardar Vallabhbhai National Institute of Technology, Surat, Gujarat-395007, India\\
	vandankp12998@gmail.com}

\author{Ajay Kumar Rai}
\address{Department of Physics, Sardar Vallabhbhai National Institute of Technology, Surat, Gujarat-395007, India\\
	raiajayk@gmail.com}

\maketitle

\begin{history}
\received{Day Month Year}
\revised{Day Month Year}
\accepted{Day Month Year}
\published{Day Month Year}
\end{history}

\begin{abstract}
We investigate the mass spectra of the \(ncnc\bar{c}\), \(nnnn\bar{c}\), and \(nnnc\bar{c}\) pentaquark systems, where \(n\) denotes a light \(u\) or \(d\) quark, within a leading-order linear Regge approximation. The extension of the baryon intercept and inverse-slope additivity relations to these systems is implemented as a minimal cluster-additive ansatz for compact color-antitriplet diquarks, rather than assumed to be an exact QCD relation. By analyzing the corresponding trajectories in the \((J,M^2)\) plane, we first determine the allowed mass range of the ground-state \(ncnc\bar{c}\) pentaquark and extract the associated Regge parameters. Using these inputs, we estimate the mass intervals of the excited states of the \(ncnc\bar{c}\), \(nnnn\bar{c}\), and \(nnnc\bar{c}\) systems. We also obtain analytical mass-inequality relations. The resulting spectra exhibit a systematic ordering with increasing total spin and orbital excitation along the Regge trajectories. We further compare the predicted states with selected kinematically compatible meson--baryon thresholds as a qualitative guide to possible nearby channels. Our results provide a useful reference for future theoretical studies and may serve as a guide for experimental searches and spin-parity assignments of pentaquark candidates in the open- and hidden-charm sectors.
\end{abstract}

\keywords{Regge Phenomenology; Pentaquark.}

\ccode{PACS numbers: 12.40.Nn, 14.20.Pt, 12.39.$-$x}

\section{Introduction}	
	In the conventional quark model, hadrons are primarily organized as mesons and baryons, corresponding to \(q\bar{q}\) and \(qqq\) configurations. Nevertheless, quantum chromodynamics does not restrict hadronic matter to these minimal quark combinations, and more complex multiquark configurations such as tetraquarks and pentaquarks are in principle allowed. 

In recent years, the study of pentaquarks has attracted renewed attention. The breakthrough in pentaquark physics came in July 2015, when the LHCb collaboration at CERN reported the first conclusive observation of pentaquark states, designated \(P_c(4450)^+\) and \(P_c(4380)^+\), in the \(\Lambda_b \to J/\psi\, p\, K^-\) decay channel \cite{fh11}. In March 2019, using a significantly larger dataset from Run 2 of the Large Hadron Collider, LHCb discovered an additional pentaquark \(P_c(4312)^+\) with a statistical significance of \(7.3\sigma\), and further analysis revealed that the previously observed \(P_c(4450)^+\) was actually composed of two distinct states, \(P_c(4440)^+\) and \(P_c(4457)^+\)\cite{fh12}.

A major milestone occurred in July 2022, when LHCb reported a narrow \(J/\psi\Lambda\) resonance, \(P_{cs}(4338)^0\), with a statistical significance above \(15\sigma\), consistent with a strange hidden-charm pentaquark candidate having the quark content \(udsc\bar{c}\)~\cite{cl1}. More broadly, the large number of exotic states discovered at the LHC, predominantly by LHCb, has contributed to what is often described as a 'particle zoo 2.0'\cite{cl3}.

Beyond the weak-decay measurements at LHCb, near-threshold charmonium production provides a complementary experimental route for investigating hidden-charm dynamics and possible pentaquark contributions. At Jefferson Lab, the near-threshold \(\gamma p\to pJ/\psi\) measurements performed by GlueX have recently been supplemented by measurements from the 007 experiment and by electroproduction data for \(\gamma^{*}p\to pJ/\psi\) from CLAS12. A combined analysis of these data provides information on the \(J/\psi p\) interaction and production mechanism, while offering complementary constraints on possible \(P_c\) contributions rather than an independent confirmation of the LHCb structures~\cite{Strakovsky2026}. Planned threshold measurements at J-PARC through the reaction \(\pi^-p\to nJ/\psi\to n(\ell^+\ell^-)\) are expected to provide an additional probe of the possible role of heavy pentaquarks in low-energy charmonium production~\cite{Strakovsky2026}.

The broader experimental program of electromagnetic meson production has also played an important role in developing the reaction and amplitude-analysis methods used in baryon spectroscopy. In particular, extensive kaon photo- and electroproduction measurements at Jefferson Lab and other facilities have generated a wide range of observables for studying strangeness-production mechanisms and excited baryons. A comprehensive account of the historical development, present experimental status, and future prospects of this program has recently been given in Ref.~\cite{Mart2026}. Although centered on strangeness production, these studies provide complementary experimental context for searches for unconventional baryonic structures.

A wide range of theoretical approaches has been employed to investigate the properties of exotic hadrons, including tetraquarks\cite{rv1,rv2,rv3} and pentaquarks\cite{17,18,19}. Pentaquark states have been investigated in a wide variety of theoretical approaches, including chromomagnetic interaction models, chiral quark models, lattice-QCD-inspired quark models, bag models, nonrelativistic potential models, and other constituent-based frameworks~\cite{20,21,22,23,24,25,26,27,28,29,30,31}. These studies show that pentaquark spectroscopy is strongly sensitive to the underlying dynamical assumptions, and no unique picture of their internal structure has yet emerged.

Recent dynamical coupled-channel studies have treated hidden-charm pentaquarks through meson--baryon scattering. Clymton, Kim, and Mart constructed the interaction kernels from effective hadronic Lagrangians and obtained the scattering amplitudes by solving a coupled Bethe--Salpeter equation using the Blankenbecler--Sugar reduction. In the nonstrange sector, which is directly relevant to the \(nnnc\bar{c}\) system considered here, channels such as \(J/\psi N\), \(\bar{D}^{(*)}\Lambda_c\), and \(\bar{D}^{(*)}\Sigma_c^{(*)}\) were included, and the pentaquark candidates were identified from resonance poles and threshold structures in the resulting amplitudes~\cite{Clymton2024}. The same coupled-channel framework has subsequently been extended to hidden-charm pentaquarks containing one and two strange quarks~\cite{Clymton2025a,Clymton2025b}. Such approaches can address pole positions, channel couplings, and resonance widths, whereas the present Regge analysis is restricted to the phenomenological organization of mass spectra and orbital excitations. The two approaches are therefore complementary and address different aspects of pentaquark spectroscopy.

Complementary reaction studies have also examined charm-sector meson--baryon channels. Skoupil and Yamaguchi studied \(\gamma p\to\bar{D}^{0}\Lambda_c^{+}\) using a Regge-plus-resonance model with Reggeized meson and baryon exchanges and \(s\)-channel \(P_c\) contributions~\cite{Skoupil2020}. Lee, Sakinah, and Oh reviewed models of \(J/\psi\) photoproduction, including Pomeron exchange within Regge phenomenology and approaches incorporating the \(J/\psi N\) interaction~\cite{Lee2022}. Sakinah, Lee, and Choi developed a dynamical \(J/\psi\) photoproduction model combining Pomeron exchange with a constituent-quark description and explicit \(J/\psi N\) final-state interactions~\cite{Sakinah2024}. These studies provide complementary dynamical context for pentaquark searches and for the phenomenological organization of mass spectra considered here.

In particular, pentaquark systems of the type \(qqqq\bar{Q}\) have long been studied in chiral constituent quark models~\cite{d1}, chromomagnetic approaches~\cite{d4}, and QCD sum rules~\cite{d7}.

More recently, several theoretical approaches, including QCD sum rules\cite{as1}, chromomagnetic interaction models \cite{as2}, and the Gürsey--Radicati mass formula \cite{as3}, have also been employed to study the dynamics of multicharm and triply heavy pentaquark systems. Such studies further highlight the diversity of possible multiquark configurations and the continuing need for systematic spectroscopy analyses in sectors where direct theoretical information remains limited.

In this context, systematic spectroscopic studies are particularly valuable, especially in flavor sectors where direct theoretical information is still limited. The \(ncnc\bar{c}\), \(nnnc\bar{c}\), and \(nnnn\bar{c}\) systems considered in the present work provide a natural laboratory for exploring charmed and multicharm pentaquark spectroscopy beyond the currently established candidates. Regge phenomenology offers a simple and efficient means of organizing hadron spectra in the \((J,M^2)\) plane and estimating excitation patterns with a limited set of inputs, making it well suited for the study of such systems. Motivated by these considerations, we investigate the mass spectra of the \(ncnc\bar{c}\), \(nnnc\bar{c}\), and \(nnnn\bar{c}\) pentaquarks within a Regge-based framework and compare the predicted states with compatible meson--baryon thresholds.

Our analysis is guided by the quasilinear Regge trajectory study of Wei \textit{et al.}~\cite{ref39}, which established a number of hadronic mass relations, including quadratic equalities and linear as well as quadratic inequalities. Adopting a similar approach with linear trajectories, we examine the correlations between slope parameters and pentaquark masses in the \((J,M^2)\) plane and derive expressions that constrain the allowed range of the ground-state mass of the \(ncnc\bar{c}\) pentaquark. We further determine the slopes and intercepts of the \((1/2)^-\), \((3/2)^-\), and \((5/2)^-\) trajectories for the \(ncnc\bar{c}\), \(nnnc\bar{c}\), and \(nnnn\bar{c}\) systems, and use these to estimate the mass intervals of the states lying on them. In this way, the trajectories and mass spectra of the three pentaquark families are constructed using one \(nnnc\bar{c}\) reference slope for each of the three spin families, without introducing additional flavor-dependent fitting parameters.

The structure of the paper is as follows. Section~2 outlines the theoretical framework of Regge phenomenology. In Sect.~3, we compute the mass spectra of the \(ncnc\bar{c}\), \(nnnn\bar{c}\), and \(nnnc\bar{c}\) pentaquark systems along the \(J^P = (1/2)^-, (3/2)^-\), and \((5/2)^-\) trajectories. The analytical mass-inequality relations used in the calculation are derived in Sect.~2. The results are discussed in Sect.~4, and the conclusions are presented in Sect.~5.

\section{Theoretical Framework}

A simple yet highly successful phenomenological framework for studying hadron spectroscopy is provided by Regge theory. In this picture, each hadron is associated with a Regge pole whose position evolves in the complex angular-momentum plane with energy. For mesons, the trajectory is characterized by whether the total spin \(J\) is even or odd, while for baryons the relevant quantity is \(J-\tfrac{1}{2}\), together with internal quantum numbers such as baryon number, parity, strangeness, charm, and bottomness, among others \cite{ref43}. The graphs of hadronic Regge trajectories in the \((J, M^2)\) plane are commonly referred to as Chew--Frautschi plots \cite{ref43,qt2}. A commonly used leading-order parametrization of a Regge trajectory in the \((J,M^2)\) plane is the linear form \cite{ru1,ru2,ru3}:

\begin{equation} \label{eq:4}
	J = \alpha(M) = \alpha(0) + \alpha' M^2,
\end{equation}

Here, \(\alpha(0)\) and \(\alpha'\) represent the intercept and slope of the corresponding particle trajectory, respectively. Hadrons that share identical internal quantum numbers and lie on a common Regge trajectory are classified as members of the same family.

In the present study, Eq.~(\ref{eq:4}) is adopted as a leading-order phenomenological ansatz and not as a statement that pentaquark Regge trajectories must be exactly linear. Recent theoretical investigations indicate that the behavior of heavy-exotic trajectories can depend on the assumed dynamics and internal clustering. Xie \emph{et al.}~\cite{Xie2024} obtained concave \(\lambda\)- and \(\rho\)-mode trajectories for hidden-bottom and hidden-charm tetraquarks, while Song \emph{et al.}~\cite{SongV2025} reported concave trajectories for the fully heavy \(P_{cc\bar{c}bb}\) pentaquark in a diquark--triquark description. In contrast, Ghasempour \emph{et al.}~\cite{Ghasempour2025} found approximately linear radial \((n_r,M^2)\) trajectories for several heavy tetraquark systems. The latter result concerns radial trajectories in a specific potential model and therefore does not directly establish linearity of the orbital \((J,M^2)\) trajectories considered here; rather, these studies collectively show that trajectory behavior remains model and excitation-mode dependent.

Introducing phenomenological curvature would require at least one additional parameter beyond the slope and intercept. Determining such a parameter requires at least three independent masses with reliable quantum-number assignments on the same trajectory. For the pentaquark sectors studied here, the presently available information provides at most two model-assigned trajectory anchors and no experimentally established sequence of states with definite \(J^P\). A nonlinear extension would consequently be underdetermined and would introduce an arbitrary curvature parameter. We therefore retain Eq.~(\ref{eq:4}) as the minimal baseline on which the subsequent Regge-additivity construction is formulated. The resulting spectra should be interpreted as predictions within this linear approximation, with possible curvature producing systematic shifts that may become more pronounced for the higher excited states.

From Eq. (\ref{eq:4}), one can derive the following expression for the slope of a hadron multiplet belonging to the same Regge trajectory.
\begin{equation} \label{eq:31}
	\alpha' = \frac{(J+1) - J}{M_{J+1}^2 - M_J^2} ,
\end{equation}

For baryon multiplets with spin-parity \(J^P\) (or, more specifically, characterized by the quantum numbers \(N^{2S+1}L_J\)) the Regge parameters, namely the slopes and intercepts, corresponding to different quark compositions can be related through the following expressions. In particular, the intercepts satisfy an additivity relation \cite{ref39,ref44,ref45,ref46,ref47,ref48,ref49,ref50,ref51,ref52}:
\begin{equation} \label{eq:5}
	\alpha_{iiq}(0) + \alpha_{jjq}(0) = 2\alpha_{ijq}(0),
\end{equation}
while the inverse slopes satisfy the additivity relation \cite{ref39,ref44,ref45,ref46,ref47}:
\begin{equation} \label{eq:6}
	\frac{1}{\alpha'_{iiq}} + \frac{1}{\alpha'_{jjq}} = \frac{2}{\alpha'_{ijq}}.
\end{equation}
Here, \(i\), \(j\), and \(q\) denote the quark flavors. Equations (\ref{eq:5}) and (\ref{eq:6}) were derived within a framework based on the topological expansion \cite{ref47}, which provides a microscopic interpretation of Regge phenomenology in terms of quark degrees of freedom \cite{ref53}. In fact, Eq. (\ref{eq:5}) was first obtained for light quarks in the dual resonance model \cite{ref48}, and was later shown to hold in the quark bremsstrahlung model \cite{ref51}, the dual-analytic model \cite{ref50}, and in two-dimensional QCD \cite{ref49}.

It is important to emphasize that Eqs.~(\ref{eq:5}) and
(\ref{eq:6}) were established for ordinary \(qqq\) baryons.
Their use in the pentaquark sector is not taken here as a
direct consequence of the baryon derivations cited above.
Instead, in Sec.~\ref{sec:cluster-additivity}, we formulate a
separate cluster-additive ansatz in which each compact
color-antitriplet diquark is treated as a single effective
constituent. Consequently, all pentaquark mass relations and
mass intervals obtained from Eqs.~(\ref{eq:5}) and
(\ref{eq:6}) should be understood as conditional predictions
within this compact-cluster approximation.

Throughout this work, and in the discussion that follows, we restrict ourselves to the ordering \(m_i \leq m_j\). This choice involves no loss of generality, since Eqs. (\ref{eq:5}) and (\ref{eq:6}) remain unchanged under the exchange \(i \leftrightarrow j\).

\subsection{\textbf{Relationship between slope ratios and masses}}

For ordinary baryons, and for an effective three-cluster
pentaquark only after imposing the cluster-additive ansatz
specified in Sec.~\ref{sec:cluster-additivity}, combining
Eqs.~(\ref{eq:4}) and (\ref{eq:5}) gives
\begin{equation} \label{eq:7}
	\alpha'_{iiq} M_{iiq}^2 + \alpha'_{jjq} M_{jjq}^2 = 2 \alpha'_{ijq} M_{ijq}^2,
\end{equation}
By combining Eqs. (\ref{eq:6}) and (\ref{eq:7}), one obtains the following quadratic equation:
\begin{equation} \label{eq:8}
	M_{jjq}^2\left( \frac{\alpha'_{jjq}}{\alpha'_{iiq}} \right)^2
	+\left( M_{iiq}^2 + M_{jjq}^2 - 4 M_{ijq}^2 \right)
	\left( \frac{\alpha'_{jjq}}{\alpha'_{iiq}} \right)
	+ M_{iiq}^2 = 0 .
\end{equation}

Upon solving the above equation, the result can be written in the following form:

\begin{equation} \label{eq:11}
	\begin{split}
		\frac{\alpha'_{jjq}}{\alpha'_{iiq}} &= \frac{1}{2 M_{jjq}^2} \biggl[ \bigl(4 M_{ijq}^2 - M_{iiq}^2 - M_{jjq}^2 \bigr) \\
		&\quad \pm \sqrt{\bigl(4 M_{ijq}^2 - M_{iiq}^2 - M_{jjq}^2 \bigr)^2 - 4 M_{iiq}^2 M_{jjq}^2} \biggr]
	\end{split},
\end{equation}


By using Eqs (\ref{eq:6}) and (\ref{eq:11}) we can get following equation for $\alpha'_{ijq}/\alpha'_{jjq}$ ,

\begin{equation}
	\label{eq:12}
	\frac{\alpha'_{ijq}}{\alpha'_{jjq}} =
	\frac{1}{4 M_{ijq}^2}
	\biggl[
	\bigl(4 M_{ijq}^2 + M_{jjq}^2 - M_{iiq}^2 \bigr)
	\pm
	\sqrt{
		\bigl(4 M_{ijq}^2 - M_{iiq}^2 - M_{jjq}^2 \bigr)^2
		- 4 M_{iiq}^2 M_{jjq}^2
	}
	\biggr].
\end{equation}

In the present analysis, we retain the solution with the positive sign in front of the square-root term. This choice is motivated by the fact that, for several well-established hadron multiplets, the corresponding slope ratios obtained from the positive branch are in much better agreement with the experimentally inferred ratios \cite{ref39}. An analogous trend is also observed for exotic hadrons: when the slope ratios are evaluated using Eq.~(\ref{eq:31}), the values are found to align more closely with the positive-root solution than with the negative-root one. Accordingly, the two equations above, with the positive sign chosen for the square-root term, can be written as

\begin{equation} \label{eq:9}
	\begin{split}
		\frac{\alpha'_{jjq}}{\alpha'_{iiq}} &= \frac{1}{2 M_{jjq}^2} \biggl[ \bigl(4 M_{ijq}^2 - M_{iiq}^2 - M_{jjq}^2 \bigr) \\
		&\quad + \sqrt{\bigl(4 M_{ijq}^2 - M_{iiq}^2 - M_{jjq}^2 \bigr)^2 - 4 M_{iiq}^2 M_{jjq}^2} \biggr]
	\end{split},
\end{equation}

\begin{equation}
	\label{eq:10}
	\frac{\alpha'_{ijq}}{\alpha'_{jjq}} =
	\frac{1}{4 M_{ijq}^2}
	\biggl[
	\bigl(4 M_{ijq}^2 + M_{jjq}^2 - M_{iiq}^2 \bigr)
	+
	\sqrt{
		\bigl(4 M_{ijq}^2 - M_{iiq}^2 - M_{jjq}^2 \bigr)^2
		- 4 M_{iiq}^2 M_{jjq}^2
	}
	\biggr].
\end{equation}

These expressions establish useful connections between the slope ratios and the masses of two-body systems. Moreover, Eq. (\ref{eq:9}) can be reformulated entirely in terms of hadron masses, thereby eliminating the explicit dependence on the slopes, by introducing an additional quark flavor \(k\). Thus, one may write

\begin{equation} \label{eq:13}
	\frac{\alpha'_{jjq}}{\alpha'_{iiq}} = \frac{\alpha'_{kkq}}{\alpha'_{iiq}} \, \frac{\alpha'_{jjq}}{\alpha'_{kkq}}.
\end{equation}

So, we will have,

\begin{equation}
	\label{eq:14}
	\begin{aligned}
		&\frac{
			(4M_{ijq}^2-M_{iiq}^2-M_{jjq}^2)
			+\sqrt{
				(4M_{ijq}^2-M_{iiq}^2-M_{jjq}^2)^2
				-4M_{iiq}^2M_{jjq}^2
			}
		}{
			2M_{jjq}^2
		}
		\\[2mm]
		&=
		\frac{
			(4M_{ikq}^2-M_{iiq}^2-M_{kkq}^2)
			+\sqrt{
				(4M_{ikq}^2-M_{iiq}^2-M_{kkq}^2)^2
				-4M_{iiq}^2M_{kkq}^2
			}
		}{
			(4M_{kjq}^2-M_{jjq}^2-M_{kkq}^2)
			+\sqrt{
				(4M_{kjq}^2-M_{jjq}^2-M_{kkq}^2)^2
				-4M_{jjq}^2M_{kkq}^2
			}
		}.
	\end{aligned}
\end{equation}

In principle, this fundamental mass relation can be used to estimate the mass of an unknown hadron state once the remaining masses are known. Nevertheless, because the pentaquark sector still lacks sufficient experimental and theoretical input, we do not make use of this relation here. Instead, we proceed by deriving a set of inequalities, as given below.

\subsection{\textbf{Linear and Quadratic Constraints on Hadron Masses}}

Two important inequalities can be derived from Eq. (\ref{eq:9}).

Because the Regge slopes \(\alpha'_{jjq}\) and \(\alpha'_{iiq}\) are required to be positive real quantities, their ratio \(\alpha'_{jjq}/\alpha'_{iiq}\) must likewise be real. It then follows from Eq. (\ref{eq:9}) that
\begin{equation} \label{eq:15}
	|4M^2_{ijq} - M^2_{iiq} - M^2_{jjq}| \geq 2M_{iiq}M_{jjq}.
\end{equation}

For the case \(i=j\), the inequality \(4M^2_{ijq} - M^2_{iiq} - M^2_{jjq} \leq 0\) is clearly not satisfied. When \(i \neq j\), this possibility is also inconsistent with the observed pattern of well-established baryon multiplets. Therefore, we infer that
\begin{equation}
	4M^2_{ijq} - M^2_{iiq} - M^2_{jjq} \geq 0.
\end{equation}  
Accordingly, Eq. (\ref{eq:15}) may be recast in the following form:

\begin{equation} \label{eq:16}
	4M^2_{ijq} - M^2_{iiq} - M^2_{jjq} \geq 2M_{iiq}M_{jjq}.
\end{equation}

Adding \(M^2_{iiq}\) and \(M^2_{jjq}\) to both sides yields

\begin{equation} \label{eq:17}
	2M_{ijq} \geq M_{iiq} + M_{jjq}.
\end{equation}

When \(i=j\), one has \(M_{iiq}=M_{ijq}=M_{jjq}\), and therefore the equality \(2M_{ijq}=M_{iiq}+M_{jjq}\) follows immediately.

On the other hand, even without imposing \(i=j\), if the relation \(2M_{ijq}=M_{iiq}+M_{jjq}\) is assumed to be valid, then Eq. (\ref{eq:9}) leads to

\begin{equation} \label{eq:18}
	\frac{\alpha'_{jjq}}{\alpha'_{iiq}} = \frac{M_{iiq}}{M_{jjq}}.
\end{equation}

From the steps leading to Eq. (\ref{eq:18}), it is clear that this relation is valid for baryons belonging to the same multiplet. Since hadrons lying on a common Regge trajectory possess the same slope, we then obtain
\begin{equation} \label{eq:19}
	\frac{\alpha'_{jjq}}{\alpha'_{iiq}} = \frac{M_{iiq,J}}{M_{jjq,J}} = \frac{M_{iiq,J+2}}{M_{jjq,J+2}}.
\end{equation}

By means of Eq. (\ref{eq:4}), one can determine the slopes associated with particular Regge trajectories. For the \(iiq\) and \(jjq\) baryons, these slopes take the form 

\begin{equation} \label{eq:20}
	\alpha'_{iiq} = \frac{2}{M^2_{iiq,J+2} - M^2_{iiq,J}}, \quad \alpha'_{jjq} = \frac{2}{M^2_{jjq,J+2} - M^2_{jjq,J}}. 
\end{equation}

So, from the above equation it follows,

\begin{equation} \label{eq:21}
	\frac{\alpha'_{jjq}}{\alpha'_{iiq}} = \frac{M_{iiq,J+2} + M_{iiq,J}}{M_{jjq,J+2} + M_{jjq,J}} \, \frac{M_{iiq,J+2} - M_{iiq,J}}{M_{jjq,J+2} - M_{jjq,J}}. 
\end{equation}

Combining Eqs. (\ref{eq:19}) and (\ref{eq:21}) gives

\begin{equation} \label{eq:22}
	\frac{\alpha'_{jjq}}{\alpha'_{iiq}} = \frac{M_{iiq,J+2} + M_{iiq,J}}{M_{jjq,J+2} + M_{jjq,J}} \, \frac{M_{iiq,J+2} - M_{iiq,J}}{M_{jjq,J+2} - M_{jjq,J}} = \left( \frac{\alpha'_{jjq}}{\alpha'_{iiq}} \right)^2. 
\end{equation}

As noted earlier, the Regge slope \(\alpha'\) must be a positive real quantity. Therefore, Eq. (\ref{eq:22}) shows that the condition \(\alpha'_{jjq}/\alpha'_{iiq}=1\) is realized when \(2M_{ijq}=M_{iiq}+M_{jjq}\). Then, from Eq. (\ref{eq:19}), it follows that \(M_{iiq,J}=M_{jjq,J}\) and \(M_{iiq,J+2}=M_{jjq,J+2}\). Within the assumed monotonic flavor-mass ordering, and excluding accidental degeneracies between states with different flavor content, the equalities \(M_{iiq,J}=M_{jjq,J}\) and \(M_{iiq,J+2}=M_{jjq,J+2}\) are identified with \(i=j\). Accordingly, under this nondegeneracy assumption, equality in Eq.~(\ref{eq:17}) occurs only for \(i=j\). Hence, for \(i\neq j\), Eq.~(\ref{eq:17}) becomes

\begin{equation} \label{eq:301}
	2M_{ijq} > M_{iiq} + M_{jjq}.
\end{equation}

By, doing internal rearrangements in the above equation we get,

\begin{equation} \label{eq:23}
	M_{jjq} < 2M_{ijq} - M_{iiq}. 
\end{equation}

A number of studies indicate that Regge-trajectory slopes tend to decrease with increasing quark mass \cite{ref44,ref45,ref100,ref101,ref102,ref103,ref104,ref47,ref53}. Consequently, when the \(j\)-quark is heavier than the \(i\)-quark, one expects \(\alpha'_{jjq}/\alpha'_{iiq} < 1\). It then follows from Eq. (\ref{eq:9}) that

\begin{equation} \label{eq:24}
	\begin{aligned}
		\frac{1}{2M_{jjq}^{2}} \Bigg[ &\left( 4M_{ijq}^{2} - M_{iiq}^{2} - M_{jjq}^{2} \right) \\
		&+ \sqrt{\left( 4M_{ijq}^{2} - M_{iiq}^{2} - M_{jjq}^{2} \right)^{2} - 4M_{iiq}^{2} M_{jjq}^{2}} \Bigg] < 1
	\end{aligned}
\end{equation}

Since the quantity under the square root in the above expression is positive, it follows that

\begin{equation} \label{eq:25}
	2M_{jjq}^2 - (4M_{ijq}^2 - M_{iiq}^2 - M_{jjq}^2) > 0
\end{equation}

By, Eqs. (\ref{eq:24}) and (\ref{eq:25}) we can have,	

\begin{equation} \label{eq:26}
	\begin{aligned}
		(4M_{ijq}^2 - M_{iiq}^2 - M_{jjq}^2)^2 
		&- 4M_{iiq}^2 M_{jjq}^2 \\
		&< \left[ 2M_{jjq}^2 - (4M_{ijq}^2 - M_{iiq}^2 - M_{jjq}^2) \right]^2
	\end{aligned}
\end{equation}

The preceding two inequalities can be reduced to the form

\begin{equation} \label{eq:27}
	M_{jjq}^2 > 2M_{ijq}^2 - M_{iiq}^2
\end{equation}

By combining Eqs. (\ref{eq:23}) and (\ref{eq:27}), one obtains the following constraint on \(M_{jjq}\).

\begin{equation} \label{eq:28}
	\sqrt{2M_{ijq}^2 - M_{iiq}^2} < M_{jjq} < 2M_{ijq} - M_{iiq}
\end{equation}

The inequality given above establishes both lower and upper limits on the mass of \(M_{jjq}\). In the next section, we will use this relation to infer the allowed mass intervals for several heavy-light pentaquark states.

	\section{Mass Spectra of Pentaquark}
\subsection{\textbf{The five-quark state in the
		diquark--diquark--antiquark picture}}
\label{sec:cluster-additivity}

In this work, we evaluate the mass spectra of the
\(ncnc\bar{c}\), \(nnnn\bar{c}\), and \(nnnc\bar{c}\)
pentaquark systems by treating them as bound states composed
of three clusters, namely two diquarks and one antiquark.
Each diquark is modeled as a correlated quark pair without
internal spatial excitation. Since a two-quark system cannot
form a color-singlet state on its own, the diquark is not
regarded as an isolated physical object, but rather as an
effective constituent that exists only inside the pentaquark.

A number of pentaquark studies employ this
diquark--diquark--antiquark reduction of the five-quark
problem. Representative examples include the
Jaffe--Wilczek-type calculation of Narodetskii
\textit{et al.}~\cite{refp2}, the potential-model treatment of
Giannuzzi~\cite{refp4}, and the effective-Hamiltonian studies
of hidden-charm pentaquarks by Shi \textit{et al.} and Ali
\textit{et al.}~\cite{refp5,re1}. These works motivate the
effective three-cluster description adopted here. They do not,
by themselves, establish the Regge-parameter additivity
relations used below.

Within this cluster description, we restrict ourselves to the
color-antitriplet diquark channel and define
\[
D_n\equiv[nn]_{\overline{\mathbf{3}}_c},
\qquad
D_c\equiv[nc]_{\overline{\mathbf{3}}_c}.
\]
Since the antiquark also transforms as
\(\overline{\mathbf{3}}_c\), a color-singlet pentaquark can be
formed through
\begin{equation}
	\overline{\mathbf{3}}_c\otimes
	\overline{\mathbf{3}}_c
	=
	\mathbf{3}_c\oplus\overline{\mathbf{6}}_c,
	\qquad
	\left[
	\left(
	\overline{\mathbf{3}}_c\otimes
	\overline{\mathbf{3}}_c
	\right)_{\mathbf{3}_c}
	\otimes\overline{\mathbf{3}}_c
	\right]_{\mathbf{1}_c}.
	\label{eq:penta-color-coupling}
\end{equation}
The color coupling of the three effective objects is therefore
the charge-conjugate analogue of the color-singlet coupling of
three quarks in an ordinary baryon. This establishes a
baryon-like three-body color topology, but it does not imply
that the dynamics of a composite diquark is identical to that
of an elementary quark.

To formulate the extension of the additivity relations
explicitly, we consider homologous states with the same
\(N^{2S+1}L_J\) assignment and the same diquark-spin coupling.
At the level of the minimal cluster-additive ansatz, the
intercept and inverse slope are written as sums of
single-cluster contributions:
\begin{align}
	\alpha_{D_aD_b\bar c}(0)
	&=
	A_{D_a}+A_{D_b}+A_{\bar c},
	\label{eq:cluster-intercept}
	\\
	\frac{1}{\alpha'_{D_aD_b\bar c}}
	&=
	B_{D_a}+B_{D_b}+B_{\bar c},
	\label{eq:cluster-inverse-slope}
\end{align}
where \(D_a,D_b\in\{D_n,D_c\}\). The quantities \(A\) and
\(B\) may depend on the Regge family under consideration; this
dependence is suppressed for notational simplicity. With the
identifications
\[
D_nD_n\bar c\equiv nnnn\bar c,\qquad
D_nD_c\bar c\equiv nnnc\bar c,\qquad
D_cD_c\bar c\equiv ncnc\bar c,
\]
Eqs.~(\ref{eq:cluster-intercept}) and
(\ref{eq:cluster-inverse-slope}) immediately give
\begin{align}
	\alpha_{nnnn\bar c}(0)+\alpha_{ncnc\bar c}(0)
	&=
	2\alpha_{nnnc\bar c}(0),
	\label{eq:penta-intercept-additivity}
	\\
	\frac{1}{\alpha'_{nnnn\bar c}}
	+
	\frac{1}{\alpha'_{ncnc\bar c}}
	&=
	\frac{2}{\alpha'_{nnnc\bar c}}.
	\label{eq:penta-slope-additivity}
\end{align}
Thus, the pentaquark relations have the same algebraic form as
the baryon relations, but here they follow from the explicitly
stated cluster-additive ansatz.

This extension assumes that the diquarks remain sufficiently
compact to be represented by effective constituents, that
their internal spatial excitations are frozen, and that the
three flavor sectors being compared correspond to homologous
spin--orbital configurations. It also neglects nonadditive
contributions associated with finite diquark size,
configuration mixing, alternative color couplings, and
genuine three-cluster interactions. Such effects would
generally produce corrections to
Eqs.~(\ref{eq:penta-intercept-additivity}) and
(\ref{eq:penta-slope-additivity}). These equations are
therefore used as phenomenological assumptions rather than
as exact five-quark identities.

The application of Eq.~(\ref{eq:28}) additionally requires an
approximately linear Regge trajectory. Approximately linear
behavior has been employed in a variety of heavy--light and
heavy--heavy systems
\cite{Ebert:2011,Chen:2018,Ebert:2011:PRD,rw9,rv10}, and Regge
parametrizations have also been applied directly to diquark
spectra~\cite{FengDiquarkRegge2023}. The latter provides
phenomenological motivation for treating a compact diquark as
an effective spectroscopic constituent, but does not
constitute a test of the additivity relations above.
Nonlinear or modified trajectories may nevertheless arise in
heavy and multiquark systems
\cite{Selem:2006,Cotugno:2009,Chen:2021,Xie2024,SongV2025}.
Because the presently available pentaquark data are
insufficient to determine additional curvature parameters,
we retain the linear trajectory as the baseline
phenomenological approximation.

Under the two stated assumptions---linear Regge behavior and
cluster additivity---the algebra leading to
Eq.~(\ref{eq:28}) remains applicable after making the
replacement
\[
i\longrightarrow D_n=[nn]_{\overline{\mathbf{3}}_c},
\qquad
j\longrightarrow D_c=[nc]_{\overline{\mathbf{3}}_c},
\qquad
q\longrightarrow\bar c.
\]
Equation~(\ref{eq:28}) can therefore be used to estimate the
pentaquark mass intervals within the adopted compact-cluster
model. The resulting inequality should not be interpreted as
a model-independent bound applicable to every possible
pentaquark configuration.

\subsection{\textbf{Mass Spectra of $nnnn\bar{c}$, \(ncnc\bar{c}\) and  $nnnc\bar{c}$ pentaquarks in the ($J,M^{2}$) plane}}

In this work we use equation (\ref{eq:28}) to evaluate the mass ranges of ground-state of the $ncnc\bar{c}$ pentaquark. $ncnc\bar{c}$  pentaquark is considered to be composed of two $nc$ diquarks and $\bar{c}$ anti-quark. Here, $n$ is up or down quark ($n=u/d$). In eq (\ref{eq:28}), if we take $i=[nn]$, $j=[nc]$, and $q=\bar{c}$ we get the following relation:

\begin{equation} \label{eq:29}
	\sqrt{2M_{nnnc\bar{c}}^2 - M_{nnnn\bar{c}}^2} < M_{ncnc\bar{c}} < 2M_{nnnc\bar{c}} - M_{nnnn\bar{c}}
\end{equation}

Experimental masses would be preferable as inputs. However, the available data do not provide the complete set of states required by the present Regge construction. No experimentally established \(nnnn\bar{c}\) pentaquark with a definite \(J^P\) is currently available. In the \(nnnc\bar{c}\) sector, several hidden-charm peak masses, including those of the \(P_c(4312)^+\), \(P_c(4440)^+\), and \(P_c(4457)^+\), have been measured, but the \(J^P\) assignments required here have not been established \cite{fh12,ref60}. The state-by-state application of Eq.~(\ref{eq:29}) requires matched masses with the same \(J^P\), while the slope relation in Eq.~(\ref{eq:31}) requires two assigned members of a common trajectory. Consequently, the experimental peak masses alone cannot be used without imposing model-dependent quantum-number and trajectory assignments.

We therefore take the required \(nnnc\bar{c}\) masses from the compact diquark model of Ref.~\cite{re1} and the \(nnnn\bar{c}\) masses from the extended chromomagnetic-interaction model of Ref.~\cite{re2}. These references are used because they provide the required states with explicit \(J^P\) assignments. We do not replace isolated entries of these spectra by experimental peak positions, since this would combine an incomplete experimental set with unmeasured, model-dependent \(J^P\) assignments. The adopted input masses are listed explicitly in Table~\ref{tab:external_mass_inputs}.

\begin{table}[ph]
	\tbl{External mass inputs used in the Regge analysis. The \(nnnn\bar{c}\) masses are taken from Ref.~\cite{re2}, whereas the \(nnnc\bar{c}\) base masses and slope partners are taken from Ref.~\cite{re1}. All masses are in GeV.\label{tab:external_mass_inputs}}
	{\footnotesize
		\setlength{\tabcolsep}{3.0pt}
		\renewcommand{\arraystretch}{1.15}
		\begin{tabular}{@{}cccc@{}} \toprule
			Trajectory &
			\(M_{nnnn\bar{c}}^{\rm base}\) &
			\(M_{nnnc\bar{c}}^{\rm base}\) &
			\(M_{nnnc\bar{c}}^{\rm partner}\) \\
			\colrule
			\(1^2S_{1/2}\rightarrow 1^2P_{3/2}\) &
			\(3.032\) &
			\(4.150\pm0.029\) &
			\(4.440\pm0.035\) \\
			
			\(1^4S_{3/2}\rightarrow 1^4P_{5/2}\) &
			\(3.043\) &
			\(4.240\pm0.029\) &
			\(4.457\pm0.035\) \\
			
			\(1^6S_{5/2}\rightarrow 1^6P_{7/2}\) &
			\(3.249\) &
			\(4.436\pm0.025\) &
			\(4.672\pm0.032\) \\
			\botrule
	\end{tabular}}
\end{table}

For the \(nnnn\bar{c}\) system, the values in Table~\ref{tab:external_mass_inputs} are the lowest \(I=1\) eigenmasses for the respective \(J^P=(1/2)^-\), \((3/2)^-\), and \((5/2)^-\) sectors in Table VIII of Ref.~\cite{re2}, rounded to the precision used in the present calculation. The uncertainties displayed for the \(nnnc\bar{c}\) masses are those quoted in Table 8 of Ref.~\cite{re1}. The values \(4.440\) and \(4.457~\mathrm{GeV}\) coincide numerically with the experimental central masses adopted in that reference; however, their \(J^P=(3/2)^+\) and \((5/2)^+\) assignments and their placement on the corresponding trajectories remain model dependent.

Ref.~\cite{re2} does not provide state-specific uncertainties or a covariance matrix for the selected \(nnnn\bar{c}\) eigenmasses. A statistically consistent propagation of all external-input uncertainties therefore cannot be constructed from the published information, and the calculations below employ the central values listed in Table~\ref{tab:external_mass_inputs}. Accordingly, the resulting mass intervals are nominal Regge bounds conditional on the adopted inputs; they are not statistical confidence intervals and inherit the model dependence of Refs.~\cite{re1,re2}.

Unless otherwise stated, the pentaquark masses derived in this work are reported to three decimal places, while the Regge slopes are quoted to three significant figures. Unrounded values are retained in all intermediate calculations.

By inserting the masses of the \(nnnc\bar{c}\) and \(nnnn\bar{c}\) states for \(J^P=(1/2)^-\), \((3/2)^-\), and \((5/2)^-\), we obtain the predicted ground-state mass intervals for the \(ncnc\bar{c}\) pentaquark. These are found to be \(5.025\text{--}5.268\) GeV for \(J^P=(1/2)^-\), \(5.167\text{--}5.437\) GeV for \(J^P=(3/2)^-\), and \(5.366\text{--}5.623\) GeV for \(J^P=(5/2)^-\), which are tabulated in Table.~\ref{tab:ncncc_pentaquark_thresholds_updated}

To estimate the higher excited states, we next evaluate the Regge slopes of these pentaquark systems. As an illustration, the slope \(\alpha'\) for the \(ncnc\bar{c}\) pentaquark can be obtained by employing relation (\ref{eq:10}). After substituting the relevant choices of \(i\) and \(j\) and rearranging the result to isolate \(\alpha'_{ncnc\bar{c}}\), we arrive at the following expression:

\begin{equation} \label{eq:30}
	\begin{split}
		\alpha'_{ncnc\bar{c}} &= \alpha'_{nnnc\bar{c}}
		\left\{
		\frac{1}{4 M_{nnnc\bar{c}}^2}
		\biggl[ \bigl(4 M_{nnnc\bar{c}}^2 + M_{ncnc\bar{c}}^2 - M_{nnnn\bar{c}}^2 \bigr) \right. \\
		&\quad \left. + \sqrt{\bigl(4 M_{nnnc\bar{c}}^2 - M_{nnnn\bar{c}}^2 - M_{ncnc\bar{c}}^2 \bigr)^2 - 4 M_{nnnn\bar{c}}^2 M_{ncnc\bar{c}}^2} \biggr]
		\right\}^{-1}
	\end{split}.
\end{equation}

At this stage, the Regge-trajectory slope for the \(nnnc\bar{c}\) pentaquark can be evaluated using Eq. (\ref{eq:31}),

\begin{equation} \label{eq:32}
	\alpha'_{nnnc\bar{c}} = \frac{1}{M_{nnnc\bar{c}(3/2)^+}^2 - M_{nnnc\bar{c}(1/2)^-}^2} .
\end{equation} 

Equation~(\ref{eq:32}) illustrates the slope extraction for the \(1^2S_{1/2}\rightarrow1^2P_{3/2}\) trajectory. For each of the three trajectories, the corresponding negative-parity base mass and positive-parity slope partner listed in Table~\ref{tab:external_mass_inputs} are used in Eq.~(\ref{eq:31}). The resulting \(nnnc\bar{c}\) Regge slopes are collected in Table~\ref{tab:Slope}. Once these slopes are obtained, Eq.~(\ref{eq:31}) is employed to determine the excited-state masses, which are listed in Table~\ref{tab:nnncc1_pentaquark_thresholds_trimmed}.

Furthermore, by inserting the values of \(M_{nnnn\bar{c}}\), \(M_{nnnc\bar{c}}\), and \(\alpha'_{nnnc\bar{c}}\) into Eq. (\ref{eq:30}), \(\alpha'_{ncnc\bar{c}}\) can be written as a function of \(M_{ncnc\bar{c}}\). Within the interval \(5.025\text{--}5.268\), this function increases monotonically. Accordingly, for \(J^P=(1/2)^-\), the allowed range of \(\alpha'_{ncnc\bar{c}}\) is found to be \(0.274\text{--}0.316\), as reported in Table \ref{tab:Slope}. The corresponding ranges for other \(J^P\) assignments are also included in Table \ref{tab:Slope}.

In an analogous way, by combining Eqs. (\ref{eq:6}) and (\ref{eq:30}), we obtain the following expression:
\\
\\



\begin{equation}
	\label{eq:33}
	\begin{aligned}
		\alpha'_{nnnn\bar{c}}
		={}&
		\alpha'_{nnnc\bar{c}}
		\Biggl[
		2
		-
		\frac{1}{4M_{nnnc\bar{c}}^2}
		\Biggl\{
		\bigl(
		4M_{nnnc\bar{c}}^2
		+
		M_{ncnc\bar{c}}^2
		-
		M_{nnnn\bar{c}}^2
		\bigr)
		\\[-1mm]
		&\quad+
		\sqrt{
			\bigl(
			4M_{nnnc\bar{c}}^2
			-
			M_{nnnn\bar{c}}^2
			-
			M_{ncnc\bar{c}}^2
			\bigr)^2
			-
			4M_{nnnn\bar{c}}^2 M_{ncnc\bar{c}}^2
		}
		\Biggr\}
		\Biggr]^{-1}.
	\end{aligned}
\end{equation}

In the above expression, substituting the values of \(M_{nnnn\bar{c}}\), \(M_{nnnc\bar{c}}\), and \(\alpha'_{nnnc\bar{c}}\) allows us to write \(\alpha'_{nnnn\bar{c}}\) as a function of \(M_{ncnc\bar{c}}\). Within the interval \(5.025\text{--}5.268\), this function decreases monotonically, yielding the range \(0.549\text{--}0.752\) for \(\alpha'_{nnnn\bar{c}}\) in the \(J^P=(1/2)^-\) case, as listed in Table \ref{tab:Slope}. The corresponding intervals of \(\alpha'_{nnnn\bar{c}}\) for the other \(J^P\) assignments are also summarized in Table \ref{tab:Slope}.

Moreover, from Eq. (\ref{eq:4}), the mass of the excited \(ncnc\bar{c}\) pentaquark state can be written as
\begin{equation} \label{eq:34}
	M_{J+m(ncnc\bar{c})} = \sqrt{M_{J(ncnc\bar{c})}^2 + \frac{m}{\alpha'_{ncnc\bar{c}}}} ,
\end{equation}
where \(m\) is an integer. By combining Eqs. (\ref{eq:30}) and (\ref{eq:34}), \(M_{J+m(ncnc\bar{c})}\) can be written as a function of \(M_{J(ncnc\bar{c})}\). As an example, for \(J^P=(1/2)^-\) and \(m=1\), evaluating this function over the interval \((5.025\text{--}5.268)\) gives a minimum value of \(5.376\) and a maximum value of \(5.563\) for \(M_{J+m(ncnc\bar{c})}\), corresponding to the \(J^P = (3/2)^+\) state. Hence, the predicted mass range for the \(ncnc\bar{c}\) pentaquark with \(J^P=(3/2)^+\) is \(5.376\text{--}5.563\), as listed in Table \ref{tab:ncncc_pentaquark_thresholds_updated}. In the same way, the higher excited states of the \(ncnc\bar{c}\) pentaquark are also evaluated and presented in Table \ref{tab:ncncc_pentaquark_thresholds_updated}.

\begin{table}[ph]
	\tbl{Values of Regge slopes for $ncnc\bar{c}$, $nnnn\bar{c}$, and $nnnc\bar{c}$ pentaquarks in the $(J,M^2)$ plane (in GeV$^{-2}$).\label{tab:Slope}}
	{\begin{tabular}{@{}cccc@{}} \toprule
			$J^P$ &
			$\alpha'_{ncnc\bar{c}}$ (GeV$^{-2}$) &
			$\alpha'_{nnnn\bar{c}}$ (GeV$^{-2}$) &
			$\alpha'_{nnnc\bar{c}}$ (GeV$^{-2}$) \\
			\colrule
			$\frac{1}{2}^{-}$ &
			0.274--0.316 &
			0.549--0.752 &
			0.401 \\
			
			$\frac{3}{2}^{-}$ &
			0.357--0.413 &
			0.738--1.030 &
			0.530 \\
			
			$\frac{5}{2}^{-}$ &
			0.318--0.367 &
			0.635--0.867 &
			0.465 \\
			\botrule
	\end{tabular}}
\end{table}

As an internal consistency check, the slope intervals in
Table~\ref{tab:Slope} satisfy
\begin{equation*}
	\alpha'_{nnnn\bar c,\mathrm{min}}
	>
	\alpha'_{nnnc\bar c}
	>
	\alpha'_{ncnc\bar c,\mathrm{max}}
	>
	0
\end{equation*}
for each of the three trajectories considered. This ordering
is physically consistent with the expected decrease of the
Regge slope when a lighter
\([nn]_{\overline{\mathbf{3}}_c}\) cluster is replaced by the
heavier \([nc]_{\overline{\mathbf{3}}_c}\) cluster. When the
correlated endpoints of the tabulated intervals are paired,
Eq.~(\ref{eq:penta-slope-additivity}) is also reproduced to
the quoted numerical precision. These observations verify the
internal implementation and the physical slope hierarchy, but
they do not constitute an independent validation of the
cluster-additive ansatz, since the slopes were obtained within
that ansatz. A direct empirical test would require homologous
Regge trajectories in all three flavor sectors, which are not
presently available.

By following the same procedure, the analogous expression for the \(nnnn\bar{c}\) pentaquark can be obtained from Eq. (\ref{eq:4}).

\begin{equation} \label{eq:35}
	M_{J+m(nnnn\bar{c})} = \sqrt{M_{J(nnnn\bar{c})}^2 + \frac{m}{\alpha'_{nnnn\bar{c}}}} ,
\end{equation}

By combining Eqs. (\ref{eq:35}) and (\ref{eq:33}), \(M_{J+m(nnnn\bar{c})}\) can be written as a function of \(M_{J(nnnn\bar{c})}\). Following this method, we have obtained the mass intervals for the remaining excited states of the \(nnnn\bar{c}\) pentaquark, and the corresponding results are listed in Table \ref{tab:nnnnc1_pentaquark_thresholds_updated}.

The predicted mass spectra of the \(ncnc\bar{c}\), \(nnnn\bar{c}\), and \(nnnc\bar{c}\) systems are given in Tables \ref{tab:ncncc_pentaquark_thresholds_updated}, \ref{tab:nnnnc1_pentaquark_thresholds_updated}, and \ref{tab:nnncc1_pentaquark_thresholds_trimmed}, respectively, as discussed above. We have also compared our results with the corresponding meson--baryon threshold values. Also in the case of \(ncnc\bar{c}\) pentaquark we have compared our results with other theoretical findings in Table.~\ref{tab:ncncc_comparison}.





The threshold masses quoted in Tables~\ref{tab:ncncc_pentaquark_thresholds_updated}, \ref{tab:nnnnc1_pentaquark_thresholds_updated}, and \ref{tab:nnncc1_pentaquark_thresholds_trimmed} are calculated as \(M_{\rm th}=m_M+m_B\), using the central hadron masses compiled by the Particle Data Group~\cite{ref60}. Only selected nearby two-body meson--baryon channels that conserve the net flavor and can couple to the indicated total \(J^P\) in relative \(S\) wave are displayed. The list is not exhaustive: other channels, including those accessible through higher relative partial waves, are not tabulated, and ``--'' means only that no nearby \(S\)-wave channel has been included. Where an isospin assignment is not specified, the channel is displayed only as a flavor- and \(J^P\)-compatible kinematic reference. For channels containing an unstable hadron, the quoted value is the nominal threshold obtained from the central masses, while the physical onset is smeared by the width of the unstable constituent. Consequently, the comparison is purely kinematical and does not by itself determine partial decay widths, coupled-channel mass shifts, binding, or the internal structure of a state.

\begin{table}[ph]
	\tbl{Mass spectra of $ncnc\bar{c}$ pentaquark states with their corresponding meson--baryon thresholds.\label{tab:ncncc_pentaquark_thresholds_updated}}
	{\setlength{\tabcolsep}{3.5pt}
		\renewcommand{\arraystretch}{1.15}
		\begin{tabular}{@{}ccccc@{}} \toprule
			State &
			$J^P$ &
			\begin{tabular}[c]{@{}c@{}}
				Calculated Mass\\
				(GeV)
			\end{tabular} &
			\begin{tabular}[c]{@{}c@{}}
				Meson--baryon\\
				thresholds
			\end{tabular} &
			\begin{tabular}[c]{@{}c@{}}
				$M_{\rm th}$\\
				(GeV)
			\end{tabular} \\
			\colrule
			
			$1^2S_{1/2}$ & $\frac{1}{2}^{-}$ & 5.025--5.268 &
			\begin{tabular}[c]{@{}c@{}}
				$\eta_c(1S)\,\Lambda_c^+$ \\
				$J/\psi(1S)\,\Lambda_c^+$
			\end{tabular} &
			\begin{tabular}[c]{@{}c@{}}
				5.270 \\
				5.383
			\end{tabular} \\
			
			$1^2P_{3/2}$ & $\frac{3}{2}^{+}$ & 5.376--5.563 &
			\begin{tabular}[c]{@{}c@{}}
				$\eta_c(1S)\,\Lambda_c(2625)^+$ \\
				$J/\psi(1S)\,\Lambda_c(2595)^+$
			\end{tabular} &
			\begin{tabular}[c]{@{}c@{}}
				5.612 \\
				5.689
			\end{tabular} \\
			
			$1^2D_{5/2}$ & $\frac{5}{2}^{-}$ & 5.706--5.848 &
			$J/\psi(1S)\,\Sigma_c(2520)^0$ &
			5.615 \\
			
			$1^2F_{7/2}$ & $\frac{7}{2}^{+}$ & 6.017--6.124 &
			$\chi_{c2}\,\Sigma_c(2520)^0$ &
			6.075 \\
			
			$1^4S_{3/2}$ & $\frac{3}{2}^{-}$ & 5.167--5.437 &
			\begin{tabular}[c]{@{}c@{}}
				$J/\psi(1S)\,\Lambda_c^+$ \\
				$\eta_c(1S)\,\Sigma_c(2520)^0$ \\
				$J/\psi(1S)\,\Sigma_c(2455)^0$
			\end{tabular} &
			\begin{tabular}[c]{@{}c@{}}
				5.383 \\
				5.502 \\
				5.551
			\end{tabular} \\
			
			$1^4P_{5/2}$ & $\frac{5}{2}^{+}$ & 5.431--5.657 &
			$J/\psi(1S)\,\Lambda_c(2625)^+$ &
			5.725 \\
			
			$1^4D_{7/2}$ & $\frac{7}{2}^{-}$ & 5.683--5.871 &
			-- &
			-- \\
			
			$1^4F_{9/2}$ & $\frac{9}{2}^{+}$ & 5.925--6.080 &
			-- &
			-- \\
			
			$1^6S_{5/2}$ & $\frac{5}{2}^{-}$ & 5.366--5.623 &
			$J/\psi(1S)\,\Sigma_c(2520)^0$ &
			5.615 \\
			
			$1^6P_{7/2}$ & $\frac{7}{2}^{+}$ & 5.652--5.862 &
			-- &
			-- \\
			
			$1^6D_{9/2}$ & $\frac{9}{2}^{-}$ & 5.924--6.095 &
			-- &
			-- \\
			
			$1^6F_{11/2}$ & $\frac{11}{2}^{+}$ & 6.184--6.322 &
			-- &
			-- \\
			
			\botrule
	\end{tabular}}
\end{table}

\begin{table}[ph]
	\tbl{Comparison of the present \(ncnc\bar{c}\) ground-state mass intervals with available theoretical \(S\)-wave results. Refs.~\cite{rm305,LiuBag2025} consider the same \(nncc\bar{c}\) quark content, whereas Refs.~\cite{rm306,rm307} concern the related \(cccn\bar{n}\) configuration. The entries from Ref.~\cite{rm305} are ordered as the \(I=0\) and \(I=1\) G\"ursey--Radicati results followed by the effective-mass result. The ranges from Ref.~\cite{LiuBag2025} include the reported \(I=0\) and \(I=1\) eigenmasses. A dagger denotes an eigenmass identified as a scattering state in the corresponding bag-model study. All masses are in GeV.\label{tab:ncncc_comparison}}
	{\scriptsize
		\setlength{\tabcolsep}{1.5pt}
		\renewcommand{\arraystretch}{1.15}
		\begin{tabular}{@{}lccccc@{}} \toprule
			\begin{tabular}[c]{@{}c@{}}
				Present state\\
				\(J^P\)
			\end{tabular} &
			\begin{tabular}[c]{@{}c@{}}
				Present\\
				work
			\end{tabular} &
			\begin{tabular}[c]{@{}c@{}}
				Ref.~\cite{rm305}\\
				\(nncc\bar{c}\)\\
				GR/effective mass
			\end{tabular} &
			\begin{tabular}[c]{@{}c@{}}
				Ref.~\cite{LiuBag2025}\\
				\(nncc\bar{c}\)\\
				MIT bag
			\end{tabular} &
			\begin{tabular}[c]{@{}c@{}}
				Ref.~\cite{rm306}\\
				\(cccn\bar{n}\)\\
				CMI
			\end{tabular} &
			\begin{tabular}[c]{@{}c@{}}
				Ref.~\cite{rm307}\\
				\(cccn\bar{n}\)\\
				MIT bag
			\end{tabular} \\
			\colrule
			
			\(1^2S_{1/2}\;(\frac{1}{2}^{-})\) &
			5.025--5.268 &
			5.71/5.78/5.84 &
			5.53--6.00 &
			5.91/6.05 &
			5.74/5.83/5.96 \\
			
			\(1^4S_{3/2}\;(\frac{3}{2}^{-})\) &
			5.167--5.437 &
			5.78/5.85/5.88 &
			5.59--5.96 &
			5.92 &
			\(5.37^{\dagger}/5.79^{\dagger}/5.85\) \\
			
			\(1^6S_{5/2}\;(\frac{5}{2}^{-})\) &
			5.366--5.623 &
			5.90/5.96/5.93 &
			\(5.84^{\dagger}/5.91^{\dagger}\) &
			-- &
			\(5.79^{\dagger}\) \\
			
			\botrule
	\end{tabular}}
\end{table}

\begin{table}[ph]
	\tbl{Mass spectra of $nnnn\bar{c}$ pentaquark states with their corresponding meson--baryon thresholds.\label{tab:nnnnc1_pentaquark_thresholds_updated}}
	{
		\setlength{\tabcolsep}{3pt}
		\renewcommand{\arraystretch}{1.15}
		\begin{tabular}{@{}ccccc@{}} \toprule
			State &
			$J^P$ &
			\begin{tabular}[c]{@{}c@{}}
				Calculated Mass\\
				(GeV)
			\end{tabular} &
			\begin{tabular}[c]{@{}c@{}}
				Meson--baryon\\
				threshold
			\end{tabular} &
			\begin{tabular}[c]{@{}c@{}}
				$M_{\rm th}$\\
				(GeV)
			\end{tabular} \\
			\colrule
			
			$1^2P_{3/2}$ & $\frac{3}{2}^{+}$ & 3.244--3.319 &
			\begin{tabular}[c]{@{}c@{}}
				$\bar{D}_1(2420)\,N$ \\
				$\bar{D}_2^{*}(2460)\,N$
			\end{tabular} &
			\begin{tabular}[c]{@{}c@{}}
				3.360 \\
				3.400
			\end{tabular} \\
			
			$1^2D_{5/2}$ & $\frac{5}{2}^{-}$ & 3.443--3.582 &
			\begin{tabular}[c]{@{}c@{}}
				$\bar{D}^{*}(2007)^0\,\Delta(1232)$ \\
				$\bar{D}^{0}\,N(1680)$
			\end{tabular} &
			\begin{tabular}[c]{@{}c@{}}
				3.239 \\
				3.550
			\end{tabular} \\
			
			$1^2F_{7/2}$ & $\frac{7}{2}^{+}$ & 3.631--3.828 &
			$\bar{D}^{*}(2007)^0\,N(1675)$ &
			3.682 \\
			
			$1^4P_{5/2}$ & $\frac{5}{2}^{+}$ & 3.199--3.258 &
			$\bar{D}_2^{*}(2460)\,N$ &
			3.400 \\
			
			$1^4D_{7/2}$ & $\frac{7}{2}^{-}$ & 3.347--3.460 &
			-- &
			-- \\
			
			$1^4F_{9/2}$ & $\frac{9}{2}^{+}$ & 3.489--3.650 &
			-- &
			-- \\
			
			$1^6P_{7/2}$ & $\frac{7}{2}^{+}$ & 3.422--3.483 &
			-- &
			-- \\
			
			$1^6D_{9/2}$ & $\frac{9}{2}^{-}$ & 3.586--3.702 &
			-- &
			-- \\
			
			$1^6F_{11/2}$ & $\frac{11}{2}^{+}$ & 3.744--3.909 &
			-- &
			-- \\
			
			\botrule
	\end{tabular}}
\end{table}

\begin{table}[ph]
	\tbl{Mass spectra of $nnnc\bar{c}$ pentaquark states with their corresponding meson--baryon thresholds.\label{tab:nnncc1_pentaquark_thresholds_trimmed}}
	{
		\setlength{\tabcolsep}{3pt}
		\renewcommand{\arraystretch}{1.15}
		\begin{tabular}{@{}ccccc@{}} \toprule
			State &
			$J^P$ &
			\begin{tabular}[c]{@{}c@{}}
				Calculated\\
				Mass\\
				(GeV)
			\end{tabular} &
			\begin{tabular}[c]{@{}c@{}}
				Meson--baryon\\
				threshold
			\end{tabular} &
			\begin{tabular}[c]{@{}c@{}}
				$M_{\rm th}$\\
				(GeV)
			\end{tabular} \\
			\colrule
			
			$1^2D_{5/2}$ & $\frac{5}{2}^{-}$ & 4.712 &
			\begin{tabular}[c]{@{}c@{}}
				$J/\psi(1S)\,\Delta(1232)$ \\
				$\eta_c(1S)\,N(1680)$ \\
				$J/\psi(1S)\,N(1680)$
			\end{tabular} &
			\begin{tabular}[c]{@{}c@{}}
				4.329 \\
				4.669 \\
				4.782
			\end{tabular} \\
			
			$1^2F_{7/2}$ & $\frac{7}{2}^{+}$ & 4.969 &
			$J/\psi(1S)\,N(1675)$ &
			4.772 \\
			
			$1^4D_{7/2}$ & $\frac{7}{2}^{-}$ & 4.664 &
			$J/\psi(1S)\,N(1680)$ &
			4.782 \\
			
			$1^4F_{9/2}$ & $\frac{9}{2}^{+}$ & 4.862 &
			-- &
			-- \\
			
			$1^6D_{9/2}$ & $\frac{9}{2}^{-}$ & 4.897 &
			$J/\psi(1S)\,\Delta(1950)$ &
			5.027 \\
			
			$1^6F_{11/2}$ & $\frac{11}{2}^{+}$ & 5.111 &
			-- &
			-- \\
			
			\botrule
	\end{tabular}}
\end{table}

	\section{Results and Discussion}

In the present work, we investigate the mass spectra of the \(ncnc\bar{c}\), \(nnnn\bar{c}\), and \(nnnc\bar{c}\) pentaquark systems within the framework of Regge phenomenology in the \((J,M^2)\) plane. We first determine the ground-state mass range of the \(ncnc\bar{c}\) pentaquark and then use the corresponding Regge framework to extract the relevant slopes for all three flavor configurations. With these inputs, we estimate the mass intervals of the excited states by constructing the associated Regge trajectories and extending them systematically to higher orbital excitations. In this way, the present analysis provides a unified description of the \(S\)- through \(F\)-wave \(ncnc\bar{c}\) states considered here, including the ground-state configurations, together with the excited states considered for the \(nnnn\bar{c}\) and \(nnnc\bar{c}\) systems.

The predicted spectra display a clear hierarchical pattern across the three flavor sectors. In particular, the \(nnnn\bar{c}\) states occupy the lowest mass region, the \(nnnc\bar{c}\) states lie at intermediate masses, and the \(ncnc\bar{c}\) states appear in the highest mass region, reflecting the expected upward shift of the spectrum with increasing charm-quark content. Within each sector, the masses increase systematically with orbital excitation, leading to a regular ordering from the lower-lying states to the higher-\(J\) excitations along the corresponding Regge trajectories in the \((J,M^2)\) plane. This overall structure indicates that the Regge-based construction provides a consistent and coherent description of the three pentaquark families and also offers a suitable basis for examining their proximity to compatible meson--baryon thresholds, which is phenomenologically important in heavy-pentaquark spectroscopy.

We first discuss the \(ncnc\bar{c}\) sector, for which the \(S\)- through \(F\)-wave states considered here, including the ground-state configurations, are obtained and compared with the available theoretical ground-state predictions.

\subsection{\texorpdfstring{\(ncnc\bar{c}\)}{ncnc cbar} pentaquark}

For the \(ncnc\bar{c}\) system, the Regge analysis yields the \(S\)- through \(F\)-wave states considered here in the \((J,M^2)\) plane.  The predicted masses extend from \(5.025\text{--}5.268~\mathrm{GeV}\)
for the \(1^2S_{1/2}\) state to
\(6.184\text{--}6.322~\mathrm{GeV}\) for the
\(1^6F_{11/2}\) state, exhibiting a regular increase with orbital
excitation and total spin. The three ground-state spin configurations, \(1^2S_{1/2}\), \(1^4S_{3/2}\), and \(1^6S_{5/2}\), are obtained in the ranges \(5.025\text{--}5.268~\mathrm{GeV}\), \(5.167\text{--}5.437~\mathrm{GeV}\), and \(5.366\text{--}5.623~\mathrm{GeV}\), respectively, while the excited \(P\)-, \(D\)-, and \(F\)-wave states follow the expected ordering along the corresponding Regge trajectories.

Table~\ref{tab:ncncc_comparison} compares the three \(S\)-wave mass intervals obtained in the present work with available theoretical results. For the same \(nncc\bar{c}\) quark content, both the G\"ursey--Radicati/effective-mass calculation of Ref.~\cite{rm305} and the MIT bag-model calculation of Ref.~\cite{LiuBag2025} give central masses above the present intervals. Depending on \(J^P\) and the model, their lowest reported masses lie approximately \(0.15\text{--}0.45~\mathrm{GeV}\) above the upper endpoints of our intervals. The \(cccn\bar{n}\) results of Refs.~\cite{rm306,rm307} involve different flavor content and are therefore included only as a qualitative multicharm comparison. The spread among these calculations demonstrates that the absolute mass scale remains strongly model dependent, even though the lowest masses generally increase with spin.

From the threshold comparison in Table~\ref{tab:ncncc_pentaquark_thresholds_updated}, several selected \(ncnc\bar{c}\) channels have nominal thresholds lying within or near the predicted mass intervals. The \(\eta_c(1S)\Lambda_c^+\) threshold lies essentially at the upper edge of the predicted \(1^2S_{1/2}\) interval. Given the numerical precision and model dependence of the calculation, no robust conclusion can be drawn as to whether this channel is open or closed, and the state should instead be regarded as exhibiting near-threshold behavior. The \(J/\psi(1S)\Lambda_c^+\) and \(J/\psi(1S)\Sigma_c(2520)^0\) thresholds associated with the \(1^4S_{3/2}\) and \(1^6S_{5/2}\) states, respectively, also lie within or near the upper portions of the corresponding predicted intervals, so their kinematic accessibility depends on the physical masses of the states. The \(1^2D_{5/2}\) interval lies entirely above the nominal \(J/\psi(1S)\Sigma_c(2520)^0\) threshold, whereas the \(\chi_{c2}(1P)\Sigma_c(2520)^0\) threshold lies within the \(1^2F_{7/2}\) interval. These comparisons identify possible nearby or kinematically accessible channels but do not determine their decay widths or imply a molecular interpretation.

\subsection{\texorpdfstring{\(nnnn\bar{c}\)}{nnnn cbar} pentaquark}

For the \(nnnn\bar{c}\) system, the Regge analysis yields the excited-state spectrum in the \((J,M^2)\) plane from the \(P\)-wave to the \(F\)-wave configurations. The predicted masses lie in the range \(3.199\text{--}3.909~\mathrm{GeV}\), with the lowest interval obtained for the \(1^4P_{5/2}\) state, \(3.199\text{--}3.258~\mathrm{GeV}\), and the highest one for the \(1^6F_{11/2}\) state, \(3.744\text{--}3.909~\mathrm{GeV}\). In general, the spectrum shows a regular increase with orbital excitation and total spin, consistent with the expected ordering of the corresponding Regge trajectories.

The threshold comparison in Table~\ref{tab:nnnnc1_pentaquark_thresholds_updated} shows that the \(1^2P_{3/2}\) interval lies below the selected \(\bar{D}_1(2420)N\) and \(\bar{D}_2^{*}(2460)N\) thresholds, while the \(1^4P_{5/2}\) interval lies below the \(\bar{D}_2^{*}(2460)N\) threshold. These particular \(S\)-wave channels are therefore nominally closed within the central-input comparison, although this does not exclude decays through other channels or higher relative partial waves. The \(1^2D_{5/2}\) interval lies above the \(\bar{D}^{*}(2007)^0\Delta(1232)\) threshold and contains the nominal \(\bar{D}^{0}N(1680)\) threshold. Similarly, the \(\bar{D}^{*}(2007)^0N(1675)\) threshold lies within the \(1^2F_{7/2}\) interval. These results establish only the kinematic locations of the selected channels; no decay-width inference is made for them or for the states whose table entries are denoted by ``--''.

Ref.~\cite{re2} provides the \(S\)-wave
	\(nnnn\bar{c}\) eigenmasses used as inputs in the present calculation,
	but does not report the corresponding \(P\)-, \(D\)-, or \(F\)-wave
	levels. Therefore, a direct state-by-state comparison is not presently
	possible, and the excited-state intervals reported here should be
	regarded as conditional Regge extrapolations from those inputs.

\subsection{\texorpdfstring{\(nnnc\bar{c}\)}{nnnc cbar} pentaquark}

For the \(nnnc\bar{c}\) system, the Regge analysis yields the excited-state spectrum in the \((J,M^2)\) plane from the \(D\)-wave to the \(F\)-wave configurations. The predicted masses lie in the range \(4.664\text{--}5.111~\mathrm{GeV}\), with the lowest mass obtained for the \(1^4D_{7/2}\) state at \(4.664~\mathrm{GeV}\) and the highest for the $1^6F_{11/2}$ state at \(5.111~\mathrm{GeV}\). In general, the spectrum shows a systematic increase with orbital excitation and total spin, consistent with the expected ordering of the corresponding Regge trajectories.

The threshold comparison in Table~\ref{tab:nnncc1_pentaquark_thresholds_trimmed} shows that the \(1^2D_{5/2}\) mass lies above the nominal \(J/\psi(1S)\Delta(1232)\) and \(\eta_c(1S)N(1680)\) thresholds, while remaining below the \(J/\psi(1S)N(1680)\) threshold. The \(1^2F_{7/2}\) mass also lies above the nominal \(J/\psi(1S)N(1675)\) threshold. The \(1^4D_{7/2}\) and \(1^6D_{9/2}\) masses lie approximately \(0.12\) and \(0.13~\mathrm{GeV}\), respectively, below the central \(J/\psi(1S)N(1680)\) and \(J/\psi(1S)\Delta(1950)\) thresholds. Since these channels contain unstable baryon resonances, their threshold comparisons should be regarded as qualitative. The listed channels therefore provide only kinematic reference points and do not determine the decay properties of the predicted states.

For comparison with a recent dynamical treatment,
	the coupled-channel calculation of Ref.~\cite{Clymton2024} obtains
	negative-parity hidden-charm poles at \(4.31\) and \(4.46~\mathrm{GeV}\)
	for \(J^P=1/2^-\), at \(4.37\), \(4.46\), and \(4.52~\mathrm{GeV}\)
	for \(J^P=3/2^-\), and at \(4.52~\mathrm{GeV}\) for \(J^P=5/2^-\).
	For the common total quantum numbers \(J^P=5/2^-\), the latter pole
	lies approximately \(0.19~\mathrm{GeV}\) below our
	\(1^2D_{5/2}\) prediction of \(4.71~\mathrm{GeV}\).
	This should not be interpreted as a state-by-state identification:
	Ref.~\cite{Clymton2024} describes the \(5/2^-\) pole predominantly
	as an \(S\)-wave \(\bar{D}^{*}\Sigma_c^{*}\) molecular state, whereas
	the present state is a \(D\)-wave excitation in the compact-cluster
	Regge construction. The coupled-channel approach determines pole
	positions, widths, and channel couplings, while the present framework
	provides a phenomenological organization of higher-orbital and
	higher-spin masses.

\section{Conclusion}

In this work, we investigated the mass spectra of the \(ncnc\bar{c}\), \(nnnn\bar{c}\), and \(nnnc\bar{c}\) pentaquark systems within the framework of Regge phenomenology in the \((J,M^2)\) plane. We first estimated the ground-state mass range of the \(ncnc\bar{c}\) pentaquark and then extracted the corresponding Regge slopes to obtain the excited-state mass intervals for all three flavor configurations. The resulting spectra exhibit a systematic increase with orbital excitation and total spin, leading to a consistent ordering of the states along the corresponding Regge trajectories. Overall, the Regge framework adopted here provides a simple and effective description of the orbital excitations of the \(ncnc\bar{c}\), \(nnnn\bar{c}\), and \(nnnc\bar{c}\) pentaquark systems using only a limited set of input parameters.

The predicted spectra are also conditional on the leading-order linear Regge ansatz. Since the available pentaquark states do not provide enough independent trajectory members to constrain a curvature parameter, no nonlinear term has been introduced. Possible deviations from linearity may shift the absolute masses, particularly for the higher orbital excitations, and can be investigated once sufficiently complete experimental trajectories with established \(J^P\) assignments become available.

Since the absolute mass scale is fixed using external theoretical spectra, the predicted mass intervals should be understood as conditional on the input assignments of Refs.~\cite{re1,re2}. Once experimentally established states with definite \(J^P\) become available in the relevant flavor sectors, these theoretical anchors can be replaced by measured masses, providing a direct experimental test and recalibration of the trajectories.

An important qualification is that the intercept and
inverse-slope additivity relations used for the pentaquarks
are not claimed to follow exactly from their original baryon
derivation. In the present analysis they represent a minimal
cluster-additive ansatz for compact color-antitriplet
diquarks in homologous spin--orbital configurations.
Accordingly, the quoted mass intervals are conditional
predictions of the linear compact-cluster Regge framework.
Finite diquark-size effects, configuration mixing,
nonadditive intercluster interactions, and coupled-channel
self-energy corrections may lead to systematic shifts that
cannot presently be quantified from the available
pentaquark data.

For the \(ncnc\bar{c}\) sector, the \(S\)- through \(F\)-wave states considered here were obtained, while for the \(nnnn\bar{c}\) and \(nnnc\bar{c}\) sectors the specified excited-state spectra were predicted. Comparison with selected nominal \(S\)-wave meson--baryon thresholds identifies several channels whose central thresholds lie within or near the predicted mass intervals, as well as some channels that are kinematically accessible in the central-input comparison. Since the tabulated channels are not exhaustive and no decay-amplitude or coupled-channel calculation is performed, these observations should be regarded only as qualitative kinematic guidance and not as predictions of stability, decay widths, binding, or molecular structure.

Comparison with the limited available theoretical results shows that the low-lying \(ncnc\bar{c}\) states obtained in the present work generally lie toward the lower side of the existing theoretical estimates, while preserving a comparable ordering of the spin configurations. Since direct theoretical information for these systems remains rather scarce, the present results provide a useful reference for future theoretical studies and may serve as a guide for experimental searches for hidden-charm pentaquark candidates in the multicharm sector. Further progress will require more detailed dynamical studies, including decay properties and coupled-channel effects, to clarify the structure of these states.

\section{Acknowledgment}
Vandan Patel gratefully acknowledges the financial support provided by the University Grants Commission (UGC) through the CSIR-UGC Junior Research Fellowship (JRF) scheme (Ref. No. 231610186052).

\end{document}